\documentclass[twocolumn,aps,prl,superscriptaddress,numerical,reprint,letterpaper,showpacs]{revtex4}
\usepackage{amstext}
\usepackage{xcolor}
\usepackage{amssymb}
\usepackage{graphicx}
\usepackage{soul}
\usepackage[utf8]{inputenc}
\usepackage{fancyhdr}
\usepackage{color}
\begin{document}
\title{Theory of Primary Photoexcitations in Donor-Acceptor Copolymers}
\author{Karan Aryanpour}
\affiliation{Department of Physics, University of Arizona, Tucson, Arizona 85721, USA}
\author{Tirthankar Dutta}
\affiliation{Department of Physics, University of Arizona, Tucson, Arizona 85721, USA}
\author{Uyen N. V. Huynh}
\affiliation{Department of Physics, University of Utah, Salt Lake City, Utah 84112, USA}
\author{Zeev Valy Vardeny}
\affiliation{Department of Physics, University of Utah, Salt Lake City, Utah 84112, USA}
\author{Sumit Mazumdar}
\email[Corresponding author.\\]{sumit@physics.arizona.edu}
\affiliation{Departments of Physics and Chemistry, University of Arizona, Tucson, Arizona 85721, USA}
\affiliation{College of Optical Sciences, University of Arizona, Tucson, Arizona 85721, USA}
\date{\today}
\begin{abstract}
We present a generic theory of primary photoexcitations in low band gap donor-acceptor conjugated copolymers.
Because of the combined effects of strong electron correlations and broken symmetry, there is considerable mixing between a charge-transfer exciton and
an energetically proximate triplet-triplet state with an overall spin singlet. The triplet-triplet state, optically forbidden
in homopolymers, is allowed in donor-acceptor copolymers. For an intermediate difference in electron affinities of the donor and the acceptor,
the triplet-triplet state can have stronger oscillator strength than the charge-transfer
exciton. We discuss the possibility of intramolecular singlet fission from the triplet-triplet state, and how such fission can be detected 
experimentally. 
\end{abstract}
\maketitle
\par The primary photophysical process in polymer solar cells is photoinduced charge transfer, whereby optical excitation at the junction between
a donor conjugated polymer and acceptor molecules 
creates a charge transfer (CT) exciton whose dissociation leads to charge carriers. The donor polymeric materials
used to be homopolymers such as polythiophene which absorb in the visible range of the solar spectrum \cite{Ma05a}.  
Homopolymers have recently been replaced by block copolymers whose repeat units consist of alternating donor ($D$) and  
acceptor ($A$) moieties \cite{Blouin07a,Chen09b,Zhang10a,Chen11a,Zhou13a,You13a,Dou13a,Kularatne13a,Hawks13a,Liu14a}. This architecture reduces the optical gap drastically, 
and the $DA$ copolymers absorb in the near infrared,
where the largest fraction of the photons emitted by the Sun lie. The power conversion efficiencies (PCEs) of organic solar cells with $DA$ copolymers as
donor materials have exceeded 10\% \cite{Liu14a}, and there is strong interest in 
the development of 
structure-property correlations that will facilitate further enhancement of the PCE. Clearly, this requires precise understanding of the nature of the
primary photoexcitations of $DA$ copolymers. 
\par Existing electronic structure calculations of $DA$ copolymers
are primarily based on the density functional theory (DFT) approach or its time-dependent version (TD-DFT) \cite{Risko11a,Risko14a,Blouin08a,Karsten09a,Pappenfus11a,OBoyle11a,Banerji12a}. 
The motivations behind these calculations have largely been to understand
the localized versus delocalized
character of the excited state reached by ground state absorption. Experimentally, $DA$ copolymers exhibit a broad low
energy (LE) absorption band at $\sim 700-800$ nm and a higher energy (HE) absorption band at $\sim 400-450$ nm \cite{Blouin07a,Chen09b,Zhang10a}. 
There is agreement between the computational studies that the LE band is due to CT from $D$ to $A$, and the HE band is a 
higher $\pi$-$\pi^{\textstyle{*}}$ excitation. 
\par Recent optical studies indicate that the above simple characterization of the LE band might be incomplete, 
and as in the homopolymers \cite{Mazumdar09a}, electron correlations play a stronger role in the photophysics of the $DA$ copolymers than envisaged
within DFT approaches.
Grancini {\it et al}. determined from ultrafast dynamics studies that the broad LE band in 
PCPDT-BT (the Supplemental Material \cite{SM} for the structures of this and other $DA$ copolymers) is composed of {\it two} distinct 
absorptions \cite{Grancini13a,Fazzi12a} centered at $725$ and $650$ nm.
TD-DFT calculations assign these to the $S_0$ $\to$ $S_1$ and $S_0$ $\to$ $S_2$ excitations, with, however, the oscillator strength of the 
second transition smaller by more than an order of magnitude \cite{Fazzi12a}. 
Two transitions underlying the LE bands in 
copolymers with CPDT as the
donor have been postulated also by Tautz {\it et al}. \cite{Tautz12a}. 
Huynh {\it et al}. have performed a transient absorption study of the $DA$ copolymer PTB7, with an optical gap $\sim 1.6$ eV 
\cite{Huynh14a}. With the pump energy at $1.55$ eV these authors found two distinct photoinduced absorptions (PAs) with the same dynamics, 
PA$_1$ at $0.4$ eV and PA$_2$ at $0.96$ eV. This is in sharp contrast to homopolymers, 
where only PA$_1$, not PA$_2$, is observed. 
Comparing against 
steady state PA measurements, Huynh {\it et al}. 
showed that (a) PA$_2$ is not a polaron absorption and (b) 
PA$_2$ overlaps strongly with PA from the lowest {\it triplet} exciton, PA$_{T_1}$ [see Figs~S2(a) and S2(b) in the Supplemental Material \cite{SM}].
These authors have obtained nearly identical results for a different $DA$ copolymer PDTP-DFBT \cite{Huynh15a}.
Busby {\it et al}. have reported 
triplet exciton generation 
in picosecond (ps) time scale from a transient absorption measurement of
the $DA$ copolymer PBTDO1 \cite{Busby15a}. The transient absorption observed is the equivalent of the higher energy PA$_2$ absorption of Huynh {\it et al}. 
\cite{Huynh14a} (see Fig.~3 in Ref.~\cite{Busby15a}). No measurement in the low energy region corresponding to PA$_1$ was reported.
The authors suggested that the triplets are generated by intramolecular singlet fission (iSF) of the optical
CT exciton.
SF is the process by which an optical
singlet exciton
dissociates into two triplet excitons
with energies half or less than that of the singlet exciton, and it is 
currently being intensively investigated as a mechanism for doubling the number of photocarriers in organic solar cells \cite{Smith13a}. 
Busby {\it et al}. noted the absence of iSF in 
PFTDO1, which has the same acceptor as PBTDO1 but a weaker donor \cite{SM}, 
in spite of the singlet and triplet energies satisfying the condition for iSF. The authors concluded that
iSF requires the strong CT character of the LE excitation \cite{Busby15a}. 
\par The above experimental results$-$in particular, the possibility 
of iSF$-$indicate that the 
theoretical treatment of $DA$ copolymers must 
incorporate electron correlation effects beyond 
TD-DFT. This is because iSF proceeds via a highly correlated
two electron-two hole ($2e$-$2h$) triplet-triplet (TT) state, which is not captured by TD-DFT 
\cite{Starcke06a,Silva-Junior08b}. Intramolecular TT states have been 
extensively discussed 
for linear polyenes, where the
lowest TT state, the $2^1A_g^-$ occurs below the optical $1^1B_u^+$ state \cite{Hudson82a}; precise description of $2e$-$2h$ states here
require configuration interaction (CI) calculations that 
include configurations quadruply excited from the 
Hartree-Fock (HF) ground state \cite{Hudson82a,Ramasesha84a,Tavan87a,Aryanpour15a}.
Unfortunately, the large and complex repeat units of the $DA$ copolymers \cite{SM} 
preclude quadruple configuration interaction (QCI) calculations and 
many-body techniques such as the density matrix 
renormalization group. Furthermore, our goal is not to explain the behavior of individual $DA$ copolymers, but rather to develop
a broad theoretical framework within which structure-property correlations may be sought. We construct here an {\it effective} 
correlated-electron theory for $DA$ copolymers that takes both of these issues into consideration.
\par Generic theoretical models of
$\pi$-conjugated homopolymers treat systems with aromatic groups or heteroatoms as
{\it ``dressed''} polyacetylenes \cite{Heeger88a,Soos89a,Soos93a}, with modified carbon (C)-atom site energies \cite{Soos89a} 
and C$-$C bond strengths \cite{Soos93a}. The goal is to understand 
low energy excitations near the optical gap. Effective theories 
miss the effects due to torsional motion of the aromatic
groups, or high energy excitations 
involving molecular orbitals (MOs) localized on the aromatic groups.
They do, however, capture the essential photophysics
near the optical gap, which is determined almost entirely by excitations from the highest valence band to the lowest
conduction band.
We adopt the same approach here. 
\begin{figure}
\includegraphics[width=3.3in]{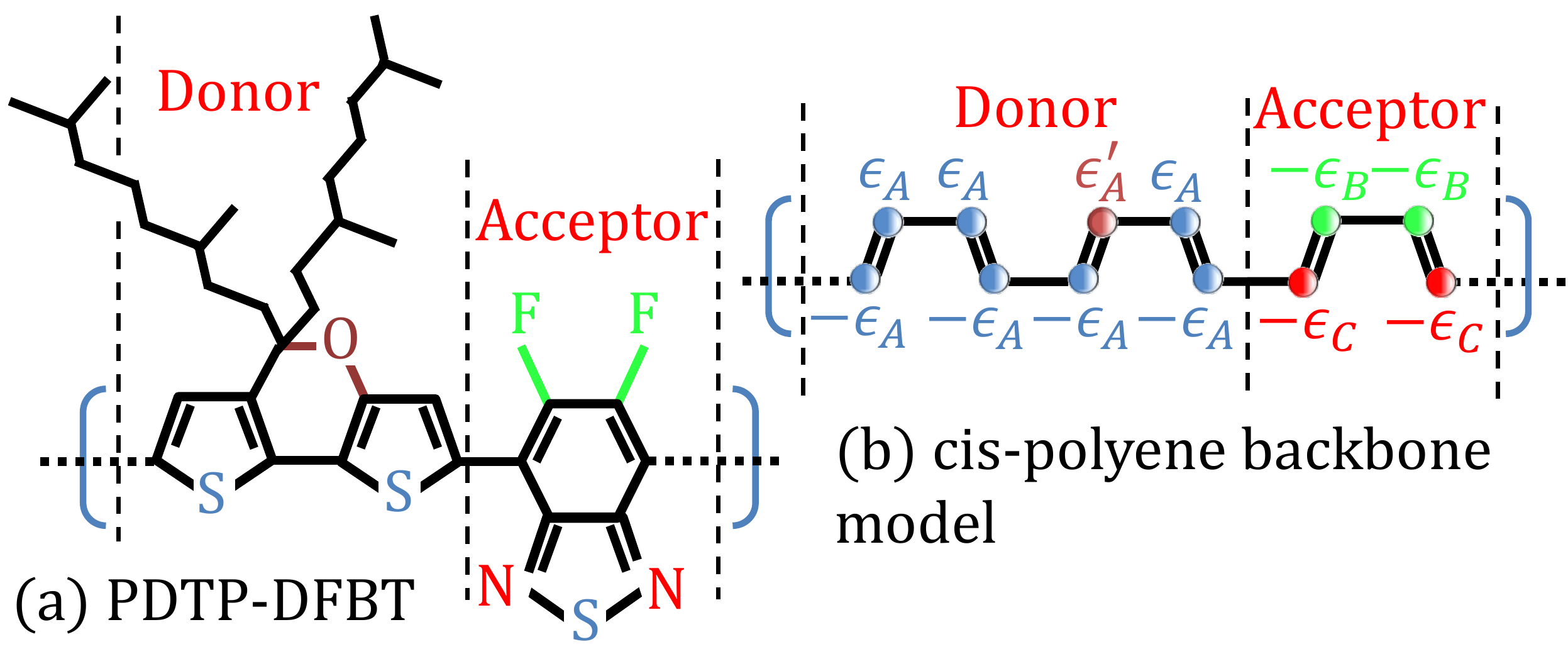}
\protect\caption{(color online). (a) PDTP-DFBT monomer. (b) 
The effective cis-polyene with the same $\pi$ conjugation path as PDTP-DFBT. The C-atom site energies reflect
the inductive effects of groups directly bonded to these atoms in PDTP-DFBT (see the text).}
\label{f1}
\end{figure}
\par We begin by developing an effective model for the $DA$ copolymer PDTP-DFBT, which when blended with PC$_{71}$BM has given the 
highest PCE in tandem solar cells \cite{You13a}. We will point out the generic nature of our theory later. The repeat unit of PDTP-DFBT is shown in
Fig.~\ref{f1}(a). The effective model cis-polyene expected to mimic the behavior of PDTP-DFBT is shown in Fig.~\ref{f1}(b). The effective polyene
has the same C$-$C $\pi$-conjugation path as the conjugated backbone of PDTP-DFBT, with the C-atom site energies 
determined by the electron affinities of the groups bonded to 
them in PDTP-DFBT. 
We investigate the monomer and
the dimer of the effective cis-polyene 
within the 
Pariser-Parr-Pople (PPP) $\pi$-electron-only
Hamiltonian \cite{Pariser53a,Pople53a},
\begin{eqnarray}
\label{PPP_Ham}
 H_{\mathrm{PPP}}=-\sum_{\langle ij \rangle\sigma}t_{ij}
(\hat{c}_{i\sigma}^{\dagger}\hat{c}_{j\sigma}^{}+\hat{c}_{j\sigma}^\dagger \hat{c}_{i\sigma}^{}) + U\sum_{i}\hat{n}_{i\uparrow} \hat{n}_{i\downarrow} \nonumber \\
 + \sum_{i<j} V_{ij} (\hat{n}_{i}-1)(\hat{n}_{j}-1)+\sum_{i}\epsilon^{}_{i}\hat{n}_{i} \,,\hspace{0.5in}
\end{eqnarray}
where $\hat{c}^{\dagger}_{i\sigma}$ creates a $\pi$ electron of spin $\sigma$ on the C atom $i$, 
$\hat{n}_{i\sigma} = \hat{c}^{\dagger}_{i\sigma}\hat{c}_{i\sigma}^{}$ is the number of electrons with spin $\sigma$ on the C atom $i$, 
$\hat{n}_{i}=\sum_{\sigma} \hat{n}_{i\sigma}$, and $\epsilon^{}_{i}$ the site energy. We use standard nearest neighbor hopping integrals
$t_{ij}=2.2$ ($2.6$) eV for single (double) C$-$C bonds. $U$ is the Coulomb repulsion between two $\pi$ electrons on the same C atom,
and $V_{ij}$ is the 
intersite Coulomb interaction. 
We parametrize the Coulomb
interactions as $V_{ij}=U/\kappa\sqrt{1+0.6117 R_{ij}^2}$, where $R_{ij}$ is the distance in angstroms between C atoms $i$ 
and $j$, and choose $U=8$ eV, $\kappa=2$ \cite{Chandross97a}. We have chosen fixed $\epsilon^{}_{A}=0.5$ eV \cite{Soos89a}
and $\epsilon_{A}^{\prime}=1.0$
eV, and 
larger $\epsilon^{}_{B}$ and $\epsilon^{}_{C}$ to reproduce the acceptor character of the DFBT group. We fix
$\epsilon^{}_{B}/\epsilon^{}_{C}=3/2$, but vary $\epsilon^{}_{B}$ to simulate the variation of the extent of CT. 
In the following, nonzero
 $\epsilon^{}_{B}$ implies that all other site energies are also nonzero. 
\par
\begin{figure}
\includegraphics[width=3.5in]{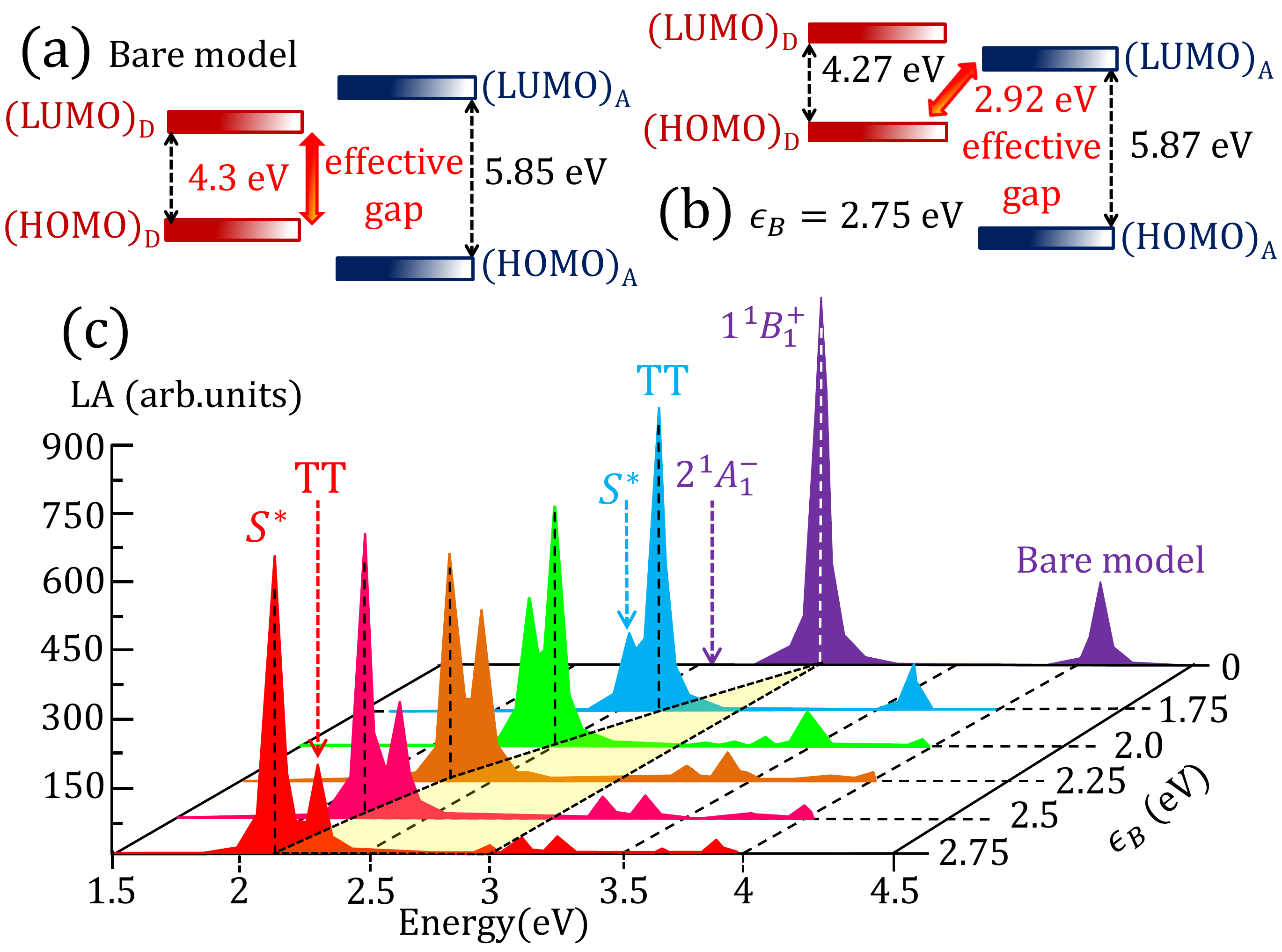}
\caption{(color online). PPP-HF HOMO and LUMO of the $D$ and $A$ segments of the monomer of Fig.~\ref{f1}(b) (a) 
for zero site energies and (b) for nonzero site energies with $\epsilon^{}_{B}=2.75$ eV. (c) Ground state absorption 
spectra of the dimer of Fig.~\ref{f1}(b) 
for a range of $\epsilon^{}_{B}$, calculated using QCI. The TT state continues to remain optically 
allowed up to $\epsilon^{}_{B}=2.75$ eV.}
\label{f2}
\end{figure}

 In Fig.~\ref{f2}(a) we have shown 
the calculated highest occupied and lowest unoccupied HF MOs (HOMOs and LUMOs) for the $D$ and $A$ groups of the ``bare'' polyene 
($\epsilon^{}_{A}=\epsilon^{\prime}_{A}=\epsilon^{}_{B}=\epsilon^{}_{C}=0$).  
Figure~\ref{f2}(b) shows the same for nonzero
site energies which reproduce the $DA$ character of the system at the HF level. 
Our calculations of ground and excited state absorptions go beyond HF, and they use exact diagonalization (full CI) for the monomer 
and QCI for the dimer of Fig.~\ref{f1}(b). 
The $C_{2v}$ and charge-conjugation symmetries of the bare polyene imply distinct one- and two-photon states, with $^1B_1^+$ and $^1A_1^-$ symmetries, 
respectively. 
Our calculated exact monomer energies of the $1^1B_1^+$ ($3.9$ eV) and $2^1A_1^-$ ($3.0$ eV) in the bare limit compare very favorably
against the experimental gas phase energies \cite{DAmico80a} of 
the $1^1B_u^+$ ($3.65$ eV) and $2^1A_g^-$ ($2.73$ eV) in trans-dodecahexaene,
allowing for the 
small differences expected between the cis- and trans-configurations, giving us confidence in our PPP parametrization.
\par Figure~\ref{f2}(c) shows our calculated QCI ground state absorption spectra for the dimer of Fig.~\ref{f1}(b) for increasing $\epsilon^{}_{B}$.
For $\epsilon^{}_{B}=0$ allowed absorption is to 1$^1B_1^+$ alone, which is of CT character. We will henceforth refer to the CT exciton as $S^{\textstyle{*}}$. 
The energy location of the dipole-forbidden $2^1A_1^-$, which is a quantum-entangled TT state with nearly
twice the energy of the lowest triplet exciton $E$($T_1$)
\cite{Hudson82a,Ramasesha84a,Tavan87a}, is indicated in the figure. For nonzero $\epsilon^{}_{B}$, 
the $C_{2v}$ symmetry 
is lost, and 
considerable configuration mixing occurs. Surprisingly, in spite of strong configuration mixing, there always exists  
a TT state at energy $\sim 2\times$E($T_1$).
The decrease in energy
of $S^{\textstyle{*}}$ with $\epsilon^{}_{B}$ is expected from the HF calculation, but the more interesting result is the 
decrease in the energy difference between $S^{\textstyle{*}}$ and TT and their crossing, when the TT is the higher energy state for $\epsilon^{}_{B} \geq 1.75$ eV. The TT  
has nonzero oscillator strength and there are two allowed absorptions.
For a range of $\epsilon^{}_{B}$ 
the two absorptions have essentially
merged, and their oscillator strengths are comparable. In the parameter range $1.75$ eV$ \leq \epsilon^{}_{B} \leq 2.125$ eV, the TT 
state actually has a larger 
oscillator strength. For still larger
$\epsilon^{}_{B}>2.25$ eV, the TT moves away from $S^{\textstyle{*}}$ and its oscillator strength begins to decrease again.
In Table~\ref{t1} we have listed the energies
of the $S^{\textstyle{*}}$ and TT states as a function of $\epsilon^{}_{B}$, for comparison against $2\times E$($T_1$). 
We will show below  that these theoretical results,
especially the intermediate coupling region, are of strong experimental relevance.
\begin{table}
{\footnotesize
\caption{QCI energies (in eV) of the two lowest singlet excited states versus twice the lowest triplet energy $E$($T_1$), for the dimer of Fig.~\ref{f1}(b), as a function of 
$\epsilon^{}_{B}$. A TT state exists for all $\epsilon^{}_{B}$. For $\epsilon^{}_{B}>1.75$ eV TT is at higher energy.} 
\label{t1}
\begin{tabular}{l}
\hline \hline
\hspace{0.8cm}$\epsilon^{}_{B}$\hspace{2.3cm}$S^{\textstyle{*}}$\hspace{2.0cm}TT\hspace{1.3cm}2$\times E$($T_1$) \\ \hline
0 (bare model)\hspace{0.8cm}$3.01$~($1^1B_1^+$)\hspace{0.8cm}$2.58$~($2^1A_1^-$)\hspace{1.0cm}$2.56$ \\
\hspace{0.9cm}1\hspace{2.2cm}$2.81$\hspace{1.9cm}$2.57$\hspace{1.5cm}$2.58$ \\
\hspace{0.7cm}1.75\hspace{2.0cm}$2.46$\hspace{1.9cm}$2.58$\hspace{1.5cm}$2.52$ \\
\hspace{0.9cm}2\hspace{2.2cm}$2.40$\hspace{1.9cm}$2.51$\hspace{1.5cm}$2.49$ \\
\hspace{0.6cm}2.125\hspace{1.95cm}$2.37$\hspace{1.9cm}$2.47$\hspace{1.5cm}$2.48$ \\
\hspace{0.7cm}2.25\hspace{2.0cm}$2.33$\hspace{1.9cm}$2.44$\hspace{1.5cm}$2.46$ \\
\hspace{0.6cm}2.375\hspace{1.95cm}$2.28$\hspace{1.9cm}$2.41$\hspace{1.5cm}$2.44$ \\
\hspace{0.8cm}2.5\hspace{2.05cm}$2.24$\hspace{1.9cm}$2.38$\hspace{1.5cm}$2.41$  \\
\hspace{0.6cm}2.625\hspace{1.95cm}$2.19$\hspace{1.9cm}$2.35$\hspace{1.5cm}$2.39$ \\
\hspace{0.7cm}2.75\hspace{2.0cm}$2.14$\hspace{1.9cm}$2.32$\hspace{1.5cm}$2.36$ \\
\hline \hline
\end{tabular}
}
\end{table}
\par
\begin{figure}
\includegraphics[width=3.4in]{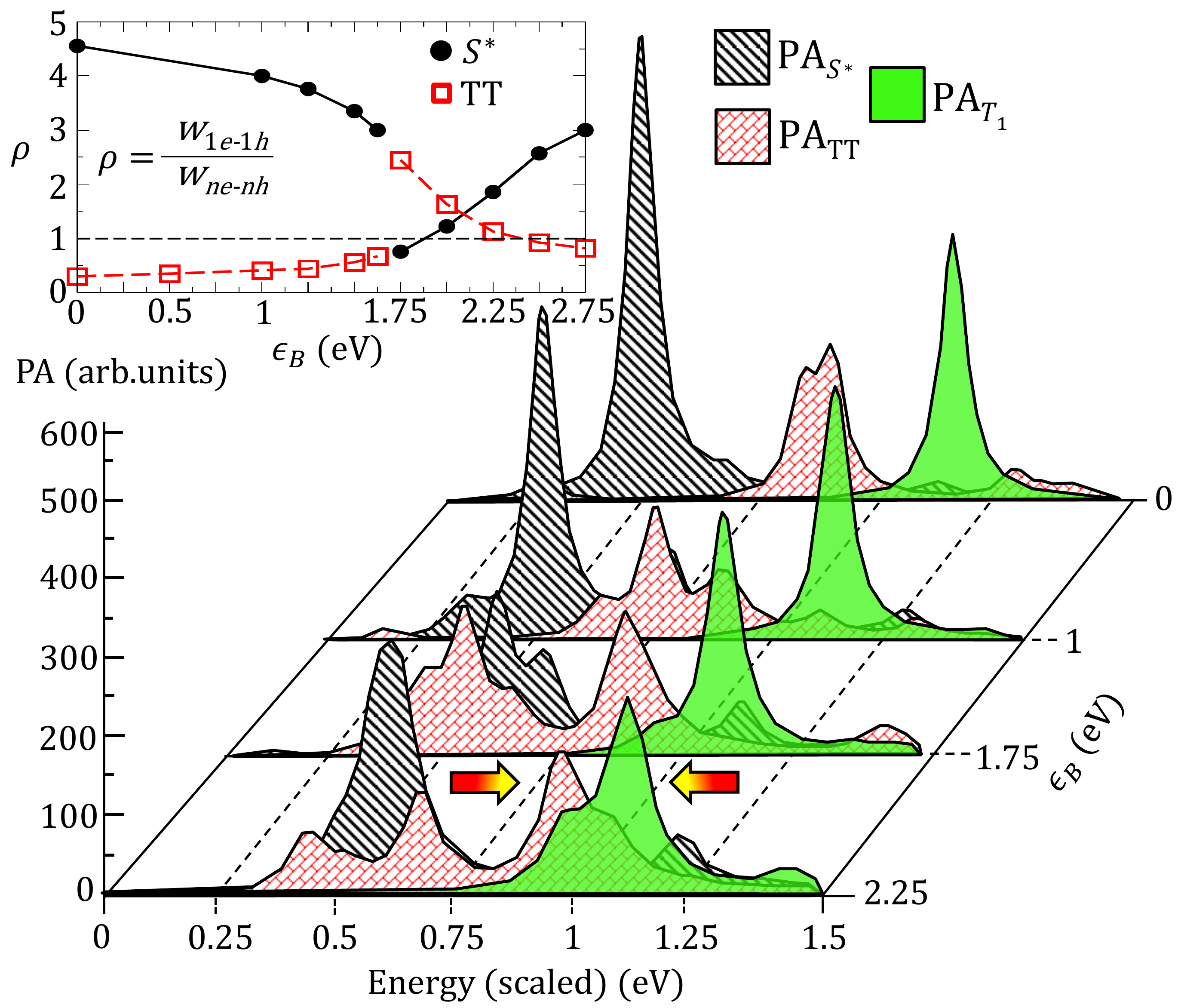}
\caption{(color online). Calculated PA$_{S^{\textstyle{*}}}$, PA$_{\mathrm{TT}}$ and PA$_{T_1}$ for the dimer of Fig.~\ref{f1}(b) for different $\epsilon^{}_{B}$'s. The arrows indicate 
a nearly complete 
overlap between the higher energy component of PA$_{\mathrm{TT}}$ and PA$_{T_1}$ at $\epsilon^{}_{B}=2.25$ eV. (Inset) The ratio of the relative weights of $1e$-$1h$ and $ne$-$nh$ ($n>1$) excitations to the QCI wave functions of $S^{\textstyle{*}}$ (circles) and 
TT (squares) states. The crossover at $\epsilon^{}_{B}=1.75$ eV is evident.}
\label{f3}
\end{figure}

 Although our calculations are for a specific dressed polyene, similar effective polyene
models can be constructed for arbitrary $DA$ copolymers. Indeed, instead of assigning multiple C-atom site energies,
a single parameter that differentiates between atoms belonging to $D$ and $A$ groups would be sufficient to derive the generic model, within which
the combined effects of electron correlations and broken symmetry 
give two optically accessible states, $S^{\textstyle{*}}$ and TT. 
We have calculated excited state absorptions from $S^{\textstyle{*}}$, TT, and $T_1$, hereafter PA$_{S^{\textstyle{*}}}$, PA$_{\mathrm{TT}}$ and PA$_{T_1}$, respectively, for the dimer of Fig.~\ref{f1}(b) to 
understand the experimental transient and steady state PA measurements \cite{Huynh14a,Huynh15a,Busby15a}. These theoretical results are shown in Fig.~\ref{f3} for several 
different $\epsilon^{}_{B}$'s. For comparison to the experimental PA spectra of different materials \cite{Huynh14a,Huynh15a,Busby15a}, 
we have normalized all PA energies by scaling against the 
optical gap of $1.55$ eV in PDTP-DFBT. 
For small $\epsilon^{}_{B} \leq 1$ eV, the calculated and experimental \cite{Huynh14a,SM} PA$_{\mathrm{TT}}$ spectra are conspicuously different. 
The calculated PA$_{\mathrm{TT}}$ and PA$_{T_1}$ bands also occur at very different energies for small $\epsilon^{}_{B}$'s.
Only, for $\epsilon^{}_{B} \geq 1.75$ eV, the calculated PA$_{\mathrm{TT}}$ resembles the experimental two-band transient 
PA$_{\mathrm{TT}}$ shown in Fig.~S2(a) of the Supplemental Material \cite{Huynh14a,Huynh15a,SM}. In the region 
$1.75$ eV $\leq \epsilon^{}_{B} \leq 2.25$ eV 
in Fig.~\ref{f2}(c),
the energy difference between $S^*$ and TT states for the dimer of Fig.~\ref{f1}(b) (corresponding to the two-unit oligomer of the PDTP-DFBT copolymer)
is negligible (see Table~\ref{t1}). 
This energy difference in the long chain limit will be vanishing relative to the C$-$C stretching frequency.
The two optical states therefore lie within the ``phonon bath'' of the copolymer and will even be coupled by electron-phonon interactions
ignored within our purely electronic model.
Thus, experimental PA$_1$ is from both states, but PA$_2$ is from TT alone (see also below). It is also worth noting that the two PA bands are 
correlated since they show the same dynamics and magnetic response \cite{Huynh15a}.
\par Quantum chemical calculations of $DA$ copolymers structurally related to PDTP-DFBT find the LUMO-LUMO offset to be nearly equal to$-$and sometimes even larger than$-$the HOMO-HOMO offset for copolymers with BT \cite{Risko11a,Risko14a,Blouin08a,Karsten09a,Pappenfus11a,OBoyle11a,Banerji12a}. We report additional calculations for the model polyene in the Supplemental Material, where the LUMO-LUMO and HOMO-HOMO offsets for the substituted polyene are nearly identical in magnitude to those reported in Ref.~\cite{Risko11a}. The results of these calculations are nearly the same as in Figs.~\ref{f2}(c) and \ref{f3}, showing very clearly that 
no generality is lost by the particular choice of MO offsets.
For each $DA$ pair,
there exist offsets where TT is optically allowed and PA$_2$ is close to PA$_{T_1}$. Conversely, two PAs, with
PA$_2$ close to PA$_{T_1}$ require that $S^{\textstyle{*}}$ and TT be nearly degenerate. PA$_1$ is from both states and PA$_2$ is from the higher energy state.
\par
\begin{figure}
\includegraphics[width=3.4in]{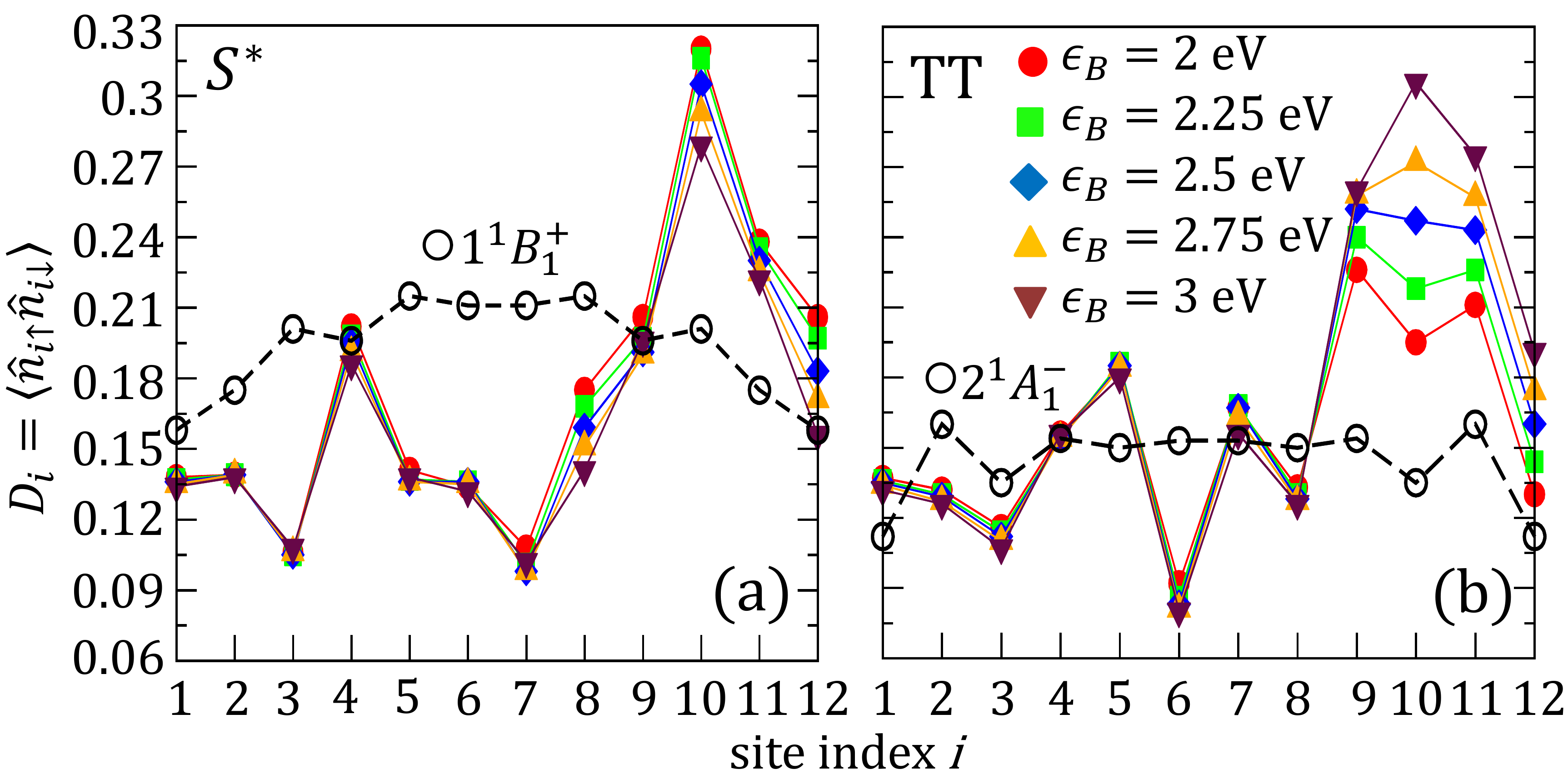}
\caption{(color online). Double occupancies by electrons of individual C-atom $p_z$ orbitals of the monomer of Fig.~\ref{f1}(b) 
for different $\epsilon^{}_{B}$'s: (a) $S^{\textstyle{*}}$, (b) TT. 
The results for 1$^1B_1^+$ and $2^1A_1^-$ states of the bare polyene are given for comparison.}
\label{f4}
\end{figure}

 $S^{\textstyle{*}}$ and TT will occur as distinct absorptions in the polymeric limit if their natures are qualitatively different.
The extent to which the wave functions of the optically allowed $S^{\textstyle{*}}$ and TT differ is therefore of interest.
The QCI excited state wave functions are 
superpositions of excitations from the HF ground state. In the bare polyene limit the $S^{\textstyle{*}}$ state is
predominantly a
$1e$-$1h$ whereas the TT has larger contributions from $ne$-$nh$ excitations ($n>1$) \cite{Silva-Junior08b,Tavan87a}.
The inset of Fig.~\ref{f3} shows the
ratio $\rho$ of the relative weights of $1e$-$1h$ versus $ne$-$nh$ excitations in the $S^{\textstyle{*}}$ and TT states as a function of $\epsilon^{}_{B}$. The 
intermediate magnitude of $\rho$ of the TT state at a moderate $\epsilon^{}_{B}$ is a signature of its partial CT character. 
In the theoretical literature, the discussion of the intramolecular TT state, the $2^1A_g^-$, has been almost entirely in the context of 
polyenes \cite{Hudson82a,Ramasesha84a,Tavan87a} or polydiacetylenes \cite{Barcza13a}. 
Within valence bond theory, the dipole-forbidden character of the $2^1A_g^-$ results from its covalent character \cite{Hudson82a,Ramasesha84a,Tavan87a}.
The ionicity of the TT versus $S^{\textstyle{*}}$ are of interest here, in view of the dipole-allowed character of the TT state.
One measure of the ionicity is $\langle n_{i,\uparrow}n_{i,\downarrow} \rangle$, the probability that the $p_z$ orbital of C atom $i$ is doubly occupied
with electrons. Exact  $\langle n_{i,\uparrow}n_{i,\downarrow} \rangle$'s for the 12-atom monomer of Fig.~\ref{f2}(b) for
both the $S^{\textstyle{*}}$ and TT states as a function of $\epsilon_{B}$ are shown in Fig.~\ref{f4}.
The asymmetry of $\langle n_{i,\uparrow}n_{i,\downarrow} \rangle$ about the chain center is indicative of the CT character of
$S^{\textstyle{*}}$. There is little change of 
$\langle n_{i,\uparrow}n_{i,\downarrow} \rangle$ in $S^{\textstyle{*}}$ for this range of $\epsilon_{B}$. In the TT state, however, $\langle n_{i,\uparrow}n_{i,\downarrow} \rangle$
increases steeply with $\epsilon_{B}$ on the C atoms constituting the acceptor  
(the C atoms constituting the $D$ group become 
positively charged, which
is not measured by $\langle n_{i,\uparrow}n_{i,\downarrow} \rangle$). {\it Covalent character is thus not a requirement for a state to be TT, as is commonly
presumed.} In addition to their ionicities, $S^{\textstyle{*}}$ and TT also differ in their bond orders, which are discussed in the Supplemental Material \cite{SM}.
\par The peculiarities noted in ultrafast spectroscopic measurements of different $DA$ copolymers \cite{Fazzi12a,Grancini13a,Huynh14a,Huynh15a,Busby15a} 
are all explained within our generic theory. 
Two close-lying ground state absorptions 
\cite{Grancini13a,Fazzi12a} and two distinct transient PA bands, with strong overlap between PA$_2$ and PA$_{T_1}$ \cite{Huynh14a,Huynh15a,SM}, 
simply require an optical TT state [see Figs.~\ref{f2}(c) and \ref{f3}], which in turn requires both strong electron correlations and broken 
spatial 
symmetry. The two peculiar observations of Busby {\it et al}. are (i) absence of triplet generation in PFTDO1 
with a weaker donor than PBTDO1 and
(ii) ultrashort lifetimes of the triplets generated
by photoexcitation: their lifetimes are 4 orders of magnitude shorter than the lifetimes of the triplets generated by sensitization. The explanations for these
observations are as follows. (i) A weak donor implies a small $\epsilon_{B}$ in Figs.~2(c) and 3; in this case
the TT state is not optically accessible and the apparent iSF is not expected. 
(ii) The short lifetimes of the triplets generated 
through photoexcitation are to be expected. Either the TT state does not undergo dissociation into individual $T_1$ at all or the partially separated $T_1$ 
pairs recombine to the TT state.
\par In summary, the photophysics of $DA$ copolymers indicate the combined effects of strong electron correlations and broken symmetry.
In the single chain limit iSF leading to complete 
separation into individual triplet excitons is unlikely, although this can occur in an aggregate or at long times.
Experimental verification of iSF would require the instrumental capability to perform transient PA experiments in the full frequency
range covering both PA$_1$ and PA$_2$: the occurrence of a single PA band$-$as opposed to two$-$would indicate iSF.
How the optically allowed character of TT in $DA$ polymers influences the PCEs of solar cells is an intriguing question and a topic for future research.
\par Work at Arizona was partially supported by NSF Grant No. DMR-1151475 and the UA-REN Faculty Exploratory Research Grant. U. H. and Z. V. acknowledge support from DOE Grant No. DE-FG02-04ER46109 and the organic semiconductors facility at the University of Utah supported by NSF-MRSEC program DMR-1121252.  
\end{document}


\title{ Supplemental Material: Theory of Primary Photoexcitations in Donor-Acceptor Copolymers}
\author{Karan Aryanpour}
\affiliation{Department of Physics, University of Arizona, Tucson, Arizona 85721, USA}
\author{Tirthankar Dutta}
\affiliation{Department of Physics, University of Arizona, Tucson, Arizona 85721, USA}
\author{Uyen N. V. Huynh}
\affiliation{Department of Physics, University of Utah, Salt Lake City, Utah 84112, USA}
\author{Zeev Valy Vardeny}
\affiliation{Department of Physics, University of Utah, Salt Lake City, Utah 84112, USA}
\author{Sumit Mazumdar}
\affiliation{Departments of Physics and Chemistry, University of Arizona, Tucson, Arizona 85721, USA}
\affiliation{College of Optical Sciences, University of Arizona, Tucson, Arizona 85721, USA}
\date{\today}
\maketitle\renewcommand{\thefigure}{S\arabic{figure}}
\renewcommand{\theequation}{S\arabic{equation}}
\section*{Structures of $DA$ copolymers discussed in the text}
\par
\begin{figure*}
\includegraphics[width=7.0in]{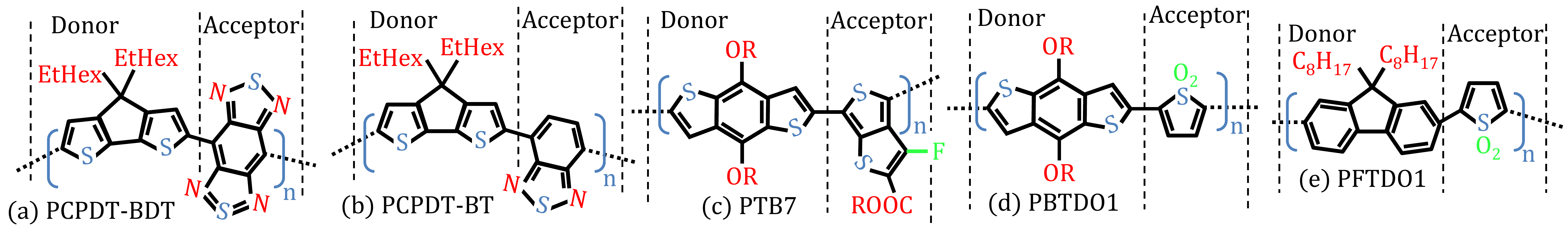}
\caption[a]{(a) PCPDT-BDT in Ref.~\cite{Tautz12a} (b) PCPDT-BT in Refs.~\cite{Tautz12a,Grancini13a} (c) PTB7 in Ref.~\cite{Huynh14a} (d) and (e) PBTDO1 and  PFTDO1, respectively in Ref.~\cite{Busby15a}}
\label{fS1}
\end{figure*}

 The structures of monomers of $DA$ copolymers other than PDTP-DFBT discussed in the text are shown in Figs.~\ref{fS1}(a)-\ref{fS1}(e). In all cases $D$ and $A$ moieties are explicitly indicated. References in which experimental work on these copolymers are reported have also been included.
\section*{Transient and steady state spectra of PTB7} 
\par The experimental setup for the transient PA measurements were discussed elsewhere \cite{Singh12b}. The transient PA measurements were done with the pump beam set at $1.55$ eV. It was composed of short pulses of $100$ fs duration, $0.1$ nJ/pulse at $80$ MHz repetition rate from a fs Ti:sapphire laser. The initial density of absorbed photons in the polymer film was lowered to be $\sim$ 10$^{16}$ cm${-3}$/pulse. The photoexcited species were monitored by PA of a probe beam with photon energy ranging from $0.55$ eV to $1.05$ eV that was generated from an OPO laser based on the Ti:sapphire laser that gives both ‘signal’ and ‘idler’ beams. In addition, we also used a differential frequency crystal (AgGa$S_2$) to generate a differential beam by phase matching of the ``signal'' and ``idler'' beams to extend the probe spectral range from $0.25$ eV to $0.43$ eV. The pump beam was modulated at frequency of $50$ kHz, and the changes, $\Delta T$ in probe transmission T were measured using a lock-in amplifier (SR830) and an InSb detector (Judson IR). The PA spectrum was obtained from the relationship PA=$-\Delta T$/$T$.
\par
\begin{figure}
\includegraphics[width=3.5in]{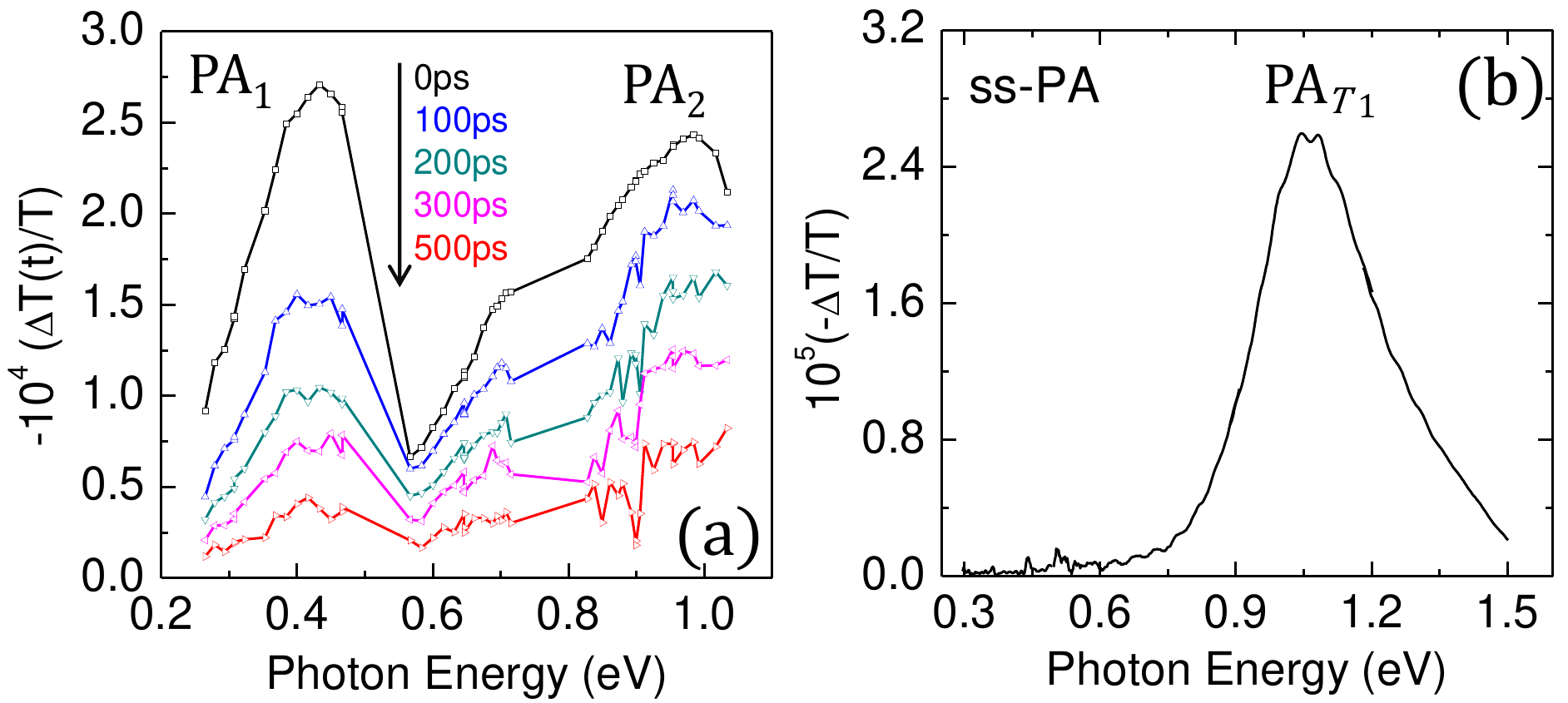}
\caption{(a) PA$_1$ and PA$_2$ (b) PA$_{T_1}$ for $DA$ copolymer PTB7 in Fig~\ref{fS1}(c).}
\label{fS2}
\end{figure}

 The steady state PA was measured using a standard experimental setup \cite{Drori07a}. Thin copolymer films were placed in a closed cycle He refrigerator cryostat at low temperature. A diode laser operating at $1.8$ photon energy of which beam was modulated at frequency, f of $310$ Hz was used as an excitation beam, and the output beam of an incandescent tungsten/halogen lamp was used as a probe light source. The changes, $\Delta T$ in the probe transmission T induced by the laser pump excitation were measured using a lock-in amplifier referenced at $f$, a monochromator, and various combinations of gratings, filters, and photodetectors spanning the spectral range $0.3 < \hbar \omega < 2.3$ eV.
In Fig.~\ref{fS2}(a) we show the time-dependent transient PA spectra for PTB7, with the pump at $1.55$ eV, which is close to the optical gap of $1.6$ eV. Two distinct PA bands, low energy PA$_1$ with a maximum at $\sim 0.4$ eV and and high energy PA$_2$ with maximum at $\sim 0.96$ eV are observed. Fig.~\ref{fS2}(b) shows the steady state PA$_{T_1}$ with maximum at $1.06$ eV. PA$_2$ occurs at the same energy region as PA$_{T_1}$. The two-band nature of the transient PA persists until $500$ ps. More recently, similar transient and steady state PA spectra have been obtained also for PDTP-DFBT in Ref.~\cite{Huynh15a}.
\section*{Bond orders of TT and $S^{\textstyle{*}}$ states: Spatial identification of triplets} 
\begin{figure}
\includegraphics[width=3.4in]{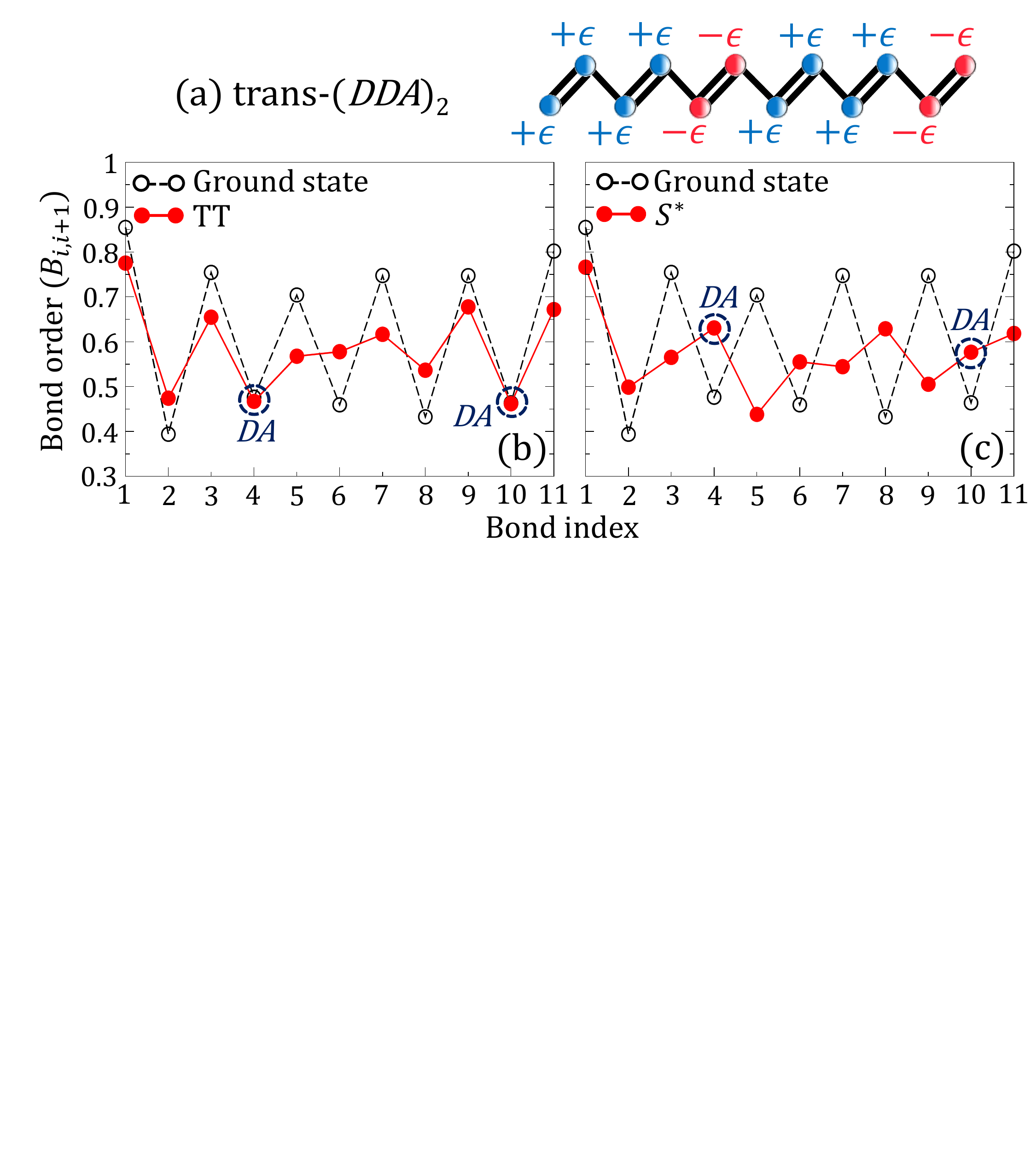}
\caption{(a) Schematic representation of trans-($DDA$)$_{2}$ showing the distribution of site energies. Sites with $+\epsilon$ ($-\epsilon$)
indicate the donor (acceptor) atoms. (b) and (c) depict the bond orders of the TT and $S^{\textstyle{*}}$
states of trans-($DDA$)$_{2}$ for $U=6$ eV, $\kappa=2$ and $\epsilon=1.75$ eV in the PPP model [Eq.~1 in main text]. Empty symbols denote the ground state and filled symbols denote
the TT (b) and $S^{\textstyle{*}}$ (c) states. Bond indices 4 and 10 in both (b) and (c) indicate $DA$ bonds within the monomers of trans-($DDA$)$_{2}$ (see text).}
\label{fS3}
\end{figure}
\par In the main text we have shown that the ``dressed'' polyene of Fig.~1(b) has two absorption bands that correspond to the states 
TT and $S^{\textstyle{*}}$, the oscillator strengths of which depend on the electron affinity difference between the $DA$ groups. 
It is however known that multiple absorptions obtained in finite system calculations can together constitute a
single band absorption in the polymer limit, except when the absorbing states (irrespective of how energetically proximate they are) 
are qualitatively different from each other. We have demonstrated that TT and $S^{\textstyle{*}}$ are indeed different based on their
electron-hole characteristics and ionicity profiles, respectively [see inset of Fig.~3 and Fig.~4 of main text]. 
Yet another quantity that distinguishes these states is the {\it bond order}, defined as, 
\begin{equation}
\label{Bond_Ord}
B_{i,i+1} = \frac{1}{2} \sum_{\sigma}\langle \hat{c}^{\dagger}_{i,\sigma}\hat{c}^{}_{i+1,\sigma} + 
\hat{c}^{\dagger}_{i+1,\sigma}\hat{c}^{}_{i,\sigma} \rangle.
\end{equation}
Here, $B_{i,i+1}$ is the bond order between nearest neighbor atoms $i$ and $i+1$, 
$\hat{c}^{\dagger}_{i,\sigma}$ ($\hat{c}^{}_{i,\sigma}$) creates (annihilates) a $\pi$-electron of spin $\sigma$ on C atom $i$, 
and $\langle \cdots \rangle$ denotes the expectation value. 
\par Apart from distinguishing the states, bond order profiles also show the spatial locations of the two 
triplets in the TT state. However, we need to consider at least a dimer for this purpose. Calculations for the 
the 24-atom dimer of Fig.~1(b) can be done only within finite order CI, which however is apt to give incorrect results 
for bond orders. We have therefore considered a different ``dressed'' polyene in which the monomer consists of 6 C atoms only. 
The ground state structure of this monomer, hereafter trans-($DDA$)$_{2}$ is shown in Fig.~\ref{fS3}(a), where the donor group 
consists of a trans-butadiene segment in which the C-atoms have positive site energies, and the acceptor group is ethylene, 
with negative site energies for the C-atoms.  We have calculated ground state absorptions and ionicities (both not shown),  
and $B_{i,i+1}$ profiles as a function of the site energies for this dimer. 
Ground state absorptions and ionicities of both trans-($DDA$)$_{2}$ and dimer of Fig.~1(b) are similar as expected. Thus, we expect 
that the bond order profiles of these two different ``dressed'' polyenes would behave similarly also. 
\par The bond order profiles of Figs.~\ref{fS3}(b) and (c) show that the TT and $S^{\textstyle{*}}$ states are different from each 
other, as well as, different from the ground state. The charge transfer exciton ($S^{\textstyle{*}}$ state) has bond orders 
(except for the bonds at the end of the chain) exactly ``out of phase'' to that of the ground state, that is, bonds that are 
weak (strong) in the ground state have become strong (weak) in the $S^{\textstyle{*}}$ state. Furthermore,  the 
$S^{\textstyle{*}}$ exciton is delocalized over almost the entire chain. On the other hand, in the TT state, all bond orders 
except those in the central region of the chain are ``in phase'' with respect to the ground state. The bonds at the chain's
center are nearly uniform indicating that the two triplets are bound and confined to this region. An important difference between 
the TT and $S^{\textstyle{*}}$ states is that, the $DA$ bonds within the monomers [bond indices 4 and 10 circled in Figs.~\ref{fS3}(b) and \ref{fS3}(c)]
 of the latter get stronger while for the former they remain 
unchanged compared to the ground state. Strengthening of the $DA$ bonds in the $S^{\textstyle{*}}$ state is a result of $D\rightarrow A$ 
charge transfer.
\section*{MO offsets between $D$ and $A$ groups versus generality of theory}
\begin{figure*}
\includegraphics[width=7.0in]{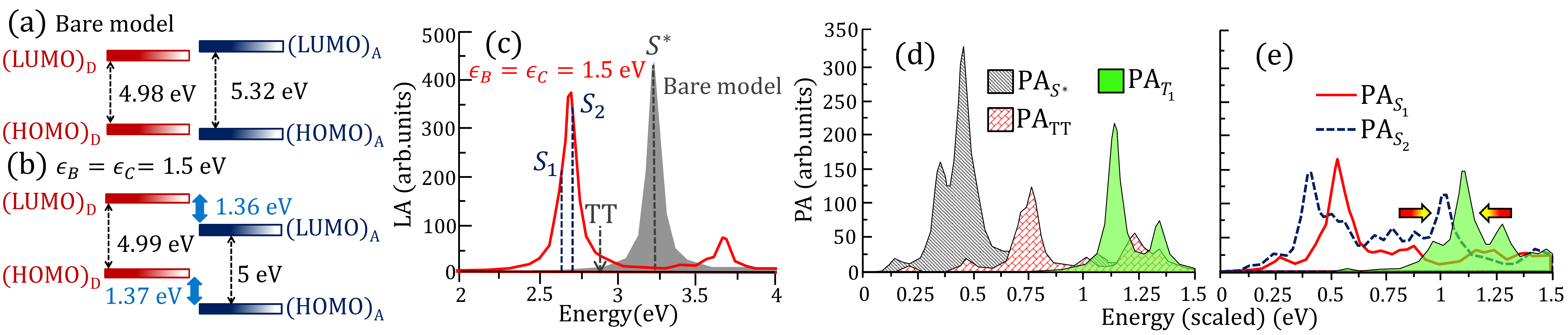}
\caption{(a) PPP-HF HOMO-LUMO gaps of the bare donor and acceptor with the modified hopping integrals (see text). (b) PPP-HF HOMOs and LUMOs of the donor and acceptor for $\epsilon^{}_{B}=\epsilon^{}_{C}=1.5$ eV. The
HOMO-HOMO and LUMO-LUMO offsets are in nearly quantitative agreement with those of Ref.~\cite{Risko11a}. (c) Calculated
QCI ground state absorptions for the bare polyene and the effective $DA$ polyene. The
squares of the transition dipole couplings of $S_1$ and $S_2$ with the ground state are indicated by the vertical dashed lines. Calculated
PAs from (d) $S^{\textstyle{*}}$, TT and $T_1$ states of the bare polyene and (e) $S_1$ and $S_2$ states of the effective $DA$ polyene.}
\label{fS4}
\end{figure*}
\par As mentioned in the main text, there exists a difference between DFT based quantum 
chemical calculations of $DA$ copolymers and our effective model polyene calculations. Within the quantum
chemical calculations of copolymers with $D$ and $A$ groups structurally related to PDTP-DFBT, the LUMO-LUMO
offsets are often larger than the HOMO-HOMO offsets, which is exactly opposite to that in Fig.~2(b) in main text.
The origin of this discrepancy lies in our choice of the donor and acceptor groups as dressed polyenes without making
any further assumption, in order to make comparisons with the bare polyene with realistic parameters. The larger length of the
substituted octatetraene representing PDTP, compared to the substituted butadiene representing the DFBT necessarily implies much smaller
HOMO-LUMO gap in the donor, ensuring smaller LUMO-LUMO offset for any choice of the site energies. In the real system the HOMO-LUMO gap
in BT is much smaller because of its extension in the direction transverse to the conjugation path. This apparent contradiction makes 
no difference whatsoever within our theory, wherein the absence of
spatial/charge conjugation symmetries and strong electron-electron interactions are the only ingredients for an optically
allowed TT state. The essential point is that PA close to experiments, with PA$_2$ overlapping with PA$_{T_1}$,
is unique to the system in which the two states reached by ground state absorption are very close in energy.
\par With the same hopping integrals, the donor
with twice as many carbon atoms as the acceptor has a HOMO-LUMO gap that is much smaller than of the acceptor,
and hence there are no site energies where the LUMO-LUMO offset is not smaller than the HOMO-HOMO offset. One way to remedy this is
to modify the $t_{ij}$ and the dimerization contribution to the HOMO-LUMO gaps within the $D$ and $A$ groups of the model polyene.
 Here we report calculations for the dimer of the effective polyene with modified hopping parameters ($t_{ij}=2.0$ and
$2.8$ eV for the donor and $t_{ij}=2.4$ eV for the acceptor), such that
the LUMO-LUMO and HOMO-HOMO offsets of the effective polyene are nearly equal, as found in Ref.~\cite{Risko11a} 
for CPDT as the donor material and BT as acceptor. The latter calculations are 
at the {\it ab initio} B3LYP/6-31G$^{\textstyle{**}}$ level of theory; the calculated HOMO-LUMO gaps for
CPDT and BT were nearly identical, and the
HOMO-HOMO (LUMO-LUMO) offsets were determined to be $1.43$ ($1.33$) eV \cite{Risko11a}. We obtain comparable
PPP-HF HOMO-LUMO gaps here (see Fig.~\ref{fS4}(a)). With site energies $\epsilon^{}_{B}=\epsilon^{}_{C}=1.5$  eV,
and with the same donor C-atom site energies as before for the polyene of Figs 2 and 3 in main text,
we obtain PPP-HF HOMO-HOMO and LUMO-LUMO offsets shown in Fig.~\ref{fS4}(b), nearly identical to those in Ref.~\cite{Risko11a}.
\par With the above site energies we then again calculated the ground state absorptions for the dimer of the effective 
polyene model of Fig.~1(b) at the QCI level. The calculated absorption spectrum
is shown in Fig.~\ref{fS4}(c), where we have also included the absorption spectrum and the location of the forbidden TT state for the
bare polyene using the same hopping parameters. Absorption in the effective $DA$ polyene is again to two states, which are now so close in energy ($\sim 0.07$ eV) that we have
not tried to distinguish between $S^{\textstyle{*}}$ and TT; rather we refer to them as $S_1$ and $S_2$. The calculated PAs from the bare polyene
(all site energies zero) are shown in Fig.~\ref{fS4}(d). As in Fig.~3 of the main text, there is no overlap between either PA$_{S^{\textstyle{*}}}$ or
PA$_{\mathrm{TT}}$ and PA$_{T_1}$ for the bare polyene. In contrast, for the $DA$ effective polyene
of Fig.~\ref{fS4}(b), we again find PA$_{S_1}$ and PA$_{S_2}$ overlapping in the low energy region, and PA$_{S_2}$ overlapping with PA$_{T_1}$
as seen in Fig.~\ref{fS4}(e). This demonstrates that independent of the HOMO-HOMO and LUMO-LUMO offsets, two distinct PAs, i.e., PA$_1$
and PA$_2$, with PA$_2$ overlapping with PA$_{T_1}$ necessarily requires optically allowed TT nearly degenerate with $S^{\textstyle{*}}$.